# EMBEDDING THE $N = 2$ SUPERSYMMETRIC YANG-MILLS THEORY IN THE ADJOINT HIGGS-YUKAWA MODEL ON THE LATTICE

I. Montvay[1]

Theoretical Physics Division, CERN
CH-1211 Geneva 23, Switzerland
and
Deutsches Elektronen-Synchrotron DESY,
Notkestr. 85, D-22603 Hamburg, Germany

## ABSTRACT

The $N = 2$ supersymmetric Yang-Mills theory is formulated on the lattice. The feasibility of numerical simulations is discussed.

[1] e-mail address: montvay@surya20.cern.ch

# 1 Introduction

Recent remarkable results in $N=2$ extended supersymmetric Yang-Mills theories [1, 2] (for further references see these papers) include exact formulas for masses and a theoretical explanation of basic phenomena such as confinement and chiral symmetry breaking. Obviously, it would be very interesting to see how these features emerge in numerical simulations. This could contribute to a better understanding of non-perturbative properties of four-dimensional supersymmetric quantum gauge field theories (for earlier reviews see [3, 4]).

A basic property of lattice regularization is that supersymmetry is broken and has to be recovered in the continuum limit. In this respect it is quite similar to another basic symmetry, namely chiral gauge symmetry [5]. In fact, the authors of ref. [6] showed that in $N=1$ supersymmetric Yang-Mills theories in the continuum limit there is an intimate connection between the restoration of supersymmetry and chiral symmetry. Following their suggestion, in the present letter I shall give a prescription on how to perform a supersymmetric continuum limit in a broader, non-supersymmetric, renormalizable quantum field theory. As an example, the $N=2$ supersymmetric SU(2) Yang-Mills theory will be considered, but a similar procedure can also be applied in other interesting theories with $N=2$ and $N=4$ extended supersymmetry.

The difficulty of considering supersymmetric theories on the lattice lies in the fact that many bare parameters are needed and the parameter tuning is a non-trivial task. In the next section the lattice action will first be defined. Then constraints implied by the convergence of the path integral will be identified and different minima of the classical potential will be classified. Section 3 is devoted to the question of how to tune the bare parameters to reach the supersymmetric fixed point in the continuum limit. Another technical difficulty comes from the doubling of fermions in popular numerical simulation algorithms. This will be discussed in the last section.

# 2 Lattice action

The fields in the $N=2$ supersymmetric SU(2) Yang-Mills theory are: $A_{x\mu}^a$, $\psi_x^i$, $\overline{\psi}_x^j$, $A_x^r$, $B_x^s$. The gauge field is $A_{x\mu}^a$, $\mu \in \{1,2,3,4\}$, $a \in \{1,2,3\}$, which is represented on the lattice by the SU(2) matrix on links $U_{x\mu} \equiv \exp(igT_a A_{x\mu}^a)$, with the SU(2) generators $T_a = \tau^a/2$. Note that for simplicity the lattice spacing (usually denoted by $a$) is set to 1 throughout this paper, in other words every dimensional quantity is measured in lattice units. The four-component fermion in the adjoint representation is described by $\psi_x^i$, $\overline{\psi}_x^j$, $i,j \in \{1,2,3\}$. The real scalar and pseudoscalar fields in the adjoint representation are denoted by $A_x^r$, $r \in \{1,2,3\}$ and $B_x^s$, $s \in \{1,2,3\}$, respectively. Instead of the component fields, sometimes, the matrices $\psi_x \equiv T_i \psi_x^i$, $\overline{\psi}_x \equiv T_j \overline{\psi}_x^j$, $A_x \equiv T_r A_x^r$, $B_x \equiv T_s B_x^s$ will also be used.

The lattice action contains three pieces

$$S = S_g + S_f + S_s . \tag{1}$$

The standard Wilson action for the gauge field $S_g$ is a sum over the plaquettes

$$S_g = \beta \sum_{pl} \left(1 - \frac{1}{2} \mathrm{Tr}\, U_{pl}\right) , \qquad (2)$$

with the bare gauge coupling given by $\beta \equiv 4/g^2$. For the fermionic part $S_f$ I shall take the Wilson formulation containing the irrelevant parameter $r$, $0 < r \leq 1$, to suppress fermion doublers in the continuum limit:

$$S_f = \sum_x \mathrm{Tr}\left\{ \mu_\psi \overline{\psi}_x \psi_x - \kappa_f \sum_{\mu=\pm 1}^{\pm 4} U_{x\mu}^\dagger \overline{\psi}_{x+\hat{\mu}} U_{x\mu}(r+\gamma_\mu)\psi_x + \overline{\psi}_x(G_A A_x + iG_B \gamma_5 B_x)\psi_x \right\} . \qquad (3)$$

Here $\mu_f$ is the bare mass parameter, $\kappa_f$ the hopping parameter, and $G_A$ and $G_B$ are the bare Yukawa couplings. $S_f$ is formulated here with Dirac fermion fields. Of course, one can rewrite it in terms of two Majorana fermion fields $\psi^{(1,2)}$ by the relation $\psi = (\psi^{(1)} + i\psi^{(2)})/\sqrt{2}$. Then the kinetic and mass terms will be multiplied by factors $\frac{1}{2}$ and the Yukawa coupling becomes off-diagonal in the Majorana index. The scalar part of the action is

$$S_s = \sum_x \mathrm{Tr}\left\{ \mu_A A_x^2 + \mu_B B_x^2 - \kappa_A \sum_{\mu=\pm 1}^{\pm 4} U_{x\mu}^\dagger A_{x+\hat{\mu}} U_{x\mu} A_x - \kappa_B \sum_{\mu=\pm 1}^{\pm 4} U_{x\mu}^\dagger B_{x+\hat{\mu}} U_{x\mu} B_x \right.$$

$$\left. + \lambda_{AAAA} A_x^4 + \lambda_{AABB} A_x^2 B_x^2 + \lambda_{BBBB} B_x^4 + \lambda_{ABAB}(A_x B_x)^2 \right\} , \qquad (4)$$

where the bare mass parameters are $\mu_{A,B}$, the hopping parameters $\kappa_{A,B}$, and $\lambda_{AAAA}, \ldots, \lambda_{ABAB}$ denote the quartic couplings. Note that in eqs. (3) and (4) the fields are in general normalization. In a numerical simulation an appropriate definition of the normalization is given by $\mu_\psi = \mu_A = \mu_B = 1$. In general, we have the following relations to the bare masses $m_\psi$, $m_A$ and $m_B$:

$$m_\psi + 4r = \frac{\mu_\psi}{2\kappa_\psi} , \qquad m_A^2 + 8 = \frac{\mu_A}{\kappa_A} , \qquad m_B^2 + 8 = \frac{\mu_B}{\kappa_B} . \qquad (5)$$

Note that the above lattice action of the SU(2) *adjoint Higgs-Yukawa model* contains all renormalizable interactions with the given set of fields if local gauge symmetry and parity conservation are required. The corresponding continuum Euclidean action is, in a conventional normalization,

$$S = \int d^4x \left\{ \frac{1}{4} F_{\mu\nu}^r(x) F_{\mu\nu}^r(x) + \frac{1}{2}(D_\mu A^r(x))(D_\mu A^r(x)) + \frac{1}{2}(D_\mu B^r(x))(D_\mu B^r(x)) \right.$$

$$+ \overline{\psi}^r(x)\gamma_\mu D_\mu \psi^r(x) + \frac{1}{2} m_A^2 A^r(x) A^r(x) + \frac{1}{2} m_B^2 B^r(x) B^r(x) + m_\psi \overline{\psi}^r(x)\psi^r(x)$$

$$+ i\epsilon_{rst}\overline{\psi}^r(x)[G_A A^s(x) + iG_B \gamma_5 B^s(x)]\psi^t(x) + \lambda_A [A^r(x) A^r(x)]^2 + \lambda_B [B^r(x) B^r(x)]^2$$

$$\left. + \lambda_{[AB]} A^r(x) A^r(x) B^s(x) B^s(x) - \lambda_{(AB)}[A^r(x) B^r(x)]^2 \right\} . \qquad (6)$$

Here $\lambda_A, \lambda_B, \lambda_{[AB]}, \lambda_{(AB)}$ are appropriate linear combinations of the quartic couplings in (4). The other notations are, as usual,

$$F_{\mu\nu}^r \equiv \partial_\mu A_\nu^r(x) - \partial_\nu A_\mu^r(x) + g\epsilon_{rst} A_\mu^s(x) A_\nu^t(x) , \qquad D_\mu A^r(x) \equiv \partial_\mu A^r(x) + g\epsilon_{rst} A_\mu^s(x) A^t(x) , \qquad (7)$$

and similarly for the covariant derivatives of the other fields.

The theory defined by the action in (6) is $N = 2$ supersymmetric if

$$m_A = m_B = m_\psi = \lambda_A = \lambda_B = 0 \ , \qquad G_A^2 = G_B^2 = 2\lambda_{[AB]} = 2\lambda_{(AB)} = g^2 \ . \tag{8}$$

(For references and a review see [7].) At zero fermion mass ($m_\psi = 0$) there is an $SU(2)_\mathcal{R} \otimes U(1)_\mathcal{A}$ global symmetry. The $SU(2)_\mathcal{R}$ part can be seen by writing the action in a (left-handed) Weyl-fermion basis by the relations

$$\psi_R(x) = C\overline{\psi}_{cL}^T \ , \qquad \overline{\psi}_R = \psi_{cL}^T C \ . \tag{9}$$

The $SU(2)_\mathcal{R}$ symmetry transformations act on the left-handed components of the fermion and antifermion fields: $(\psi_L, \psi_{cL})$. The $U(1)_\mathcal{A}$ symmetry is the usual singlet axial symmetry acting on fermions and augmented by a chiral rotation of the scalar fields:

$$\psi(x)' = e^{-i\alpha\gamma_5}\psi(x) \ , \qquad \overline{\psi}(x)' = \overline{\psi}(x)e^{-i\alpha\gamma_5} \ ,$$

$$A(x)' = \cos(2\alpha)A(x) - \sin(2\alpha)B(x) \ , \qquad B(x)' = \sin(2\alpha)A(x) + \cos(2\alpha)B(x) \ . \tag{10}$$

This turns out to be an anomalous classical symmetry of the action in (6) if

$$m_\psi = 0 \ , \qquad m_A^2 = m_B^2 \ , \qquad G_A = G_B \ , \qquad 2\lambda_A = 2\lambda_B = \lambda_{[AB]} - \lambda_{(AB)} \ . \tag{11}$$

The fermion mass term breaks down $SU(2)_\mathcal{R}$ to its diagonal subgroup $U(1)_\mathcal{F}$, which corresponds to the fermion number conservation. In the lattice formulation the Wilson term proportional to $r$ breaks $SU(2)_\mathcal{R} \otimes U(1)_\mathcal{A}$ to $U(1)_\mathcal{F}$ even at zero bare fermion mass $m_\psi = 0$. The rest of $SU(2)_\mathcal{R}$ has to be restored in the massless continuum limit, similarly to non-singlet axial symmetries with Wilson fermions.

In order that the path integral over the scalar fields with the lattice action (1)-(4) be convergent, the quartic couplings dominating at large fields have to fulfil the following conditions:

$$\lambda_A > 0 \quad \text{AND} \quad \lambda_B > 0 \quad \text{AND}$$

$$\left\{ \lambda_{[AB]} \geq \max(0, \lambda_{(AB)}) \quad \text{OR} \quad 4\lambda_A\lambda_B > \max[\lambda_{[AB]}^2, (\lambda_{[AB]} - \lambda_{(AB)})^2] \right\} \ . \tag{12}$$

The positivity of $\lambda_{A,B}$ is in conflict with the supersymmetry condition in eq. (8). Therefore at finite lattice spacing one has to take $\lambda_{A,B} > 0$ and tune them in the continuum limit to zero (see next section). Note that in principle one could stabilize the path integral by adding irrelevant higher-dimensional terms to the action, e.g. $(A^r A^r)^3$. However, this would only be a complication and could not solve the problem because in the continuum limit the instability would reappear.

The phase structure of the lattice model is, of course, the first important question for non-perturbative studies. As a hint for possible phases one considers the minima of the classical potential

$$V(A, B) \equiv \frac{1}{2}m_A^2 A^r A^r + \frac{1}{2}m_B^2 B^r B^r + \lambda_A(A^r A^r)^2 + \lambda_B(B^r B^r)^2 + \lambda_{[AB]} A^r A^r B^s B^s - \lambda_{(AB)}(A^r B^r)^2 \ . \tag{13}$$

This has the following extrema, with the corresponding values of $V$,

$$A^r = B^r = 0 \ , \qquad V_{min} = 0 \ ; \tag{14}$$

$$A^r = 0 \ , \ B^r B^r = -\frac{m_B^2}{4\lambda_B} \ , \qquad V_{min} = -\frac{m_B^4}{16\lambda_B} \ ; \tag{15}$$

$$B^r = 0 \ , \ A^r A^r = -\frac{m_A^2}{4\lambda_A} \ , \qquad V_{min} = -\frac{m_A^4}{16\lambda_A} \ ; \tag{16}$$

$$A^r B^r = 0 \ , \ A^r A^r = \frac{m_A^2 \lambda_B - \frac{1}{2} m_B^2 \lambda_{[AB]}}{\lambda_{[AB]}^2 - 4\lambda_A \lambda_B} \ , \ B^r B^r = \frac{m_B^2 \lambda_A - \frac{1}{2} m_A^2 \lambda_{[AB]}}{\lambda_{[AB]}^2 - 4\lambda_A \lambda_B} \ ,$$

$$V_{min} = \frac{-m_A^2 m_B^2 \lambda_{[AB]} + m_A^4 \lambda_B + m_B^4 \lambda_A}{4(\lambda_{[AB]}^2 - 4\lambda_A \lambda_B)} \ ; \tag{17}$$

$$(A^r B^r)^2 = A^r A^r B^s B^s \ ,$$

$$A^r A^r = \frac{m_A^2 \lambda_B - \frac{1}{2} m_B^2 (\lambda_{[AB]} - \lambda_{(AB)})}{(\lambda_{[AB]} - \lambda_{(AB)})^2 - 4\lambda_A \lambda_B} \ , \ B^r B^r = \frac{m_B^2 \lambda_A - \frac{1}{2} m_A^2 (\lambda_{[AB]} - \lambda_{(AB)})}{(\lambda_{[AB]} - \lambda_{(AB)})^2 - 4\lambda_A \lambda_B} \ ,$$

$$V_{min} = \frac{-m_A^2 m_B^2 (\lambda_{[AB]} - \lambda_{(AB)}) + m_A^4 \lambda_B + m_B^4 \lambda_A}{4[(\lambda_{[AB]} - \lambda_{(AB)})^2 - 4\lambda_A \lambda_B]} \ . \tag{18}$$

The first minimum in (14) corresponds to the symmetric phase. All the others are defining phases with the Higgs mechanism and occur for negative scalar bare mass squares $m_A^2, m_B^2 < 0$. The one in (17) breaks the SU(2) gauge symmetry completely and will presumably not play any rôle in supersymmetry. It will not be considered in what follows. The others break the SU(2) gauge symmetry to U(1). The last one in (18) gives two degenerate minima with $A$ and $B$ either parallel or antiparallel to each other and, of course, assumes that the denominators are non-vanishing. Otherwise, in the general case, there is no corresponding minimum at all, except if the bare masses satisfy the condition

$$4\lambda_A \lambda_B = (\lambda_{[AB]} - \lambda_{(AB)})^2 \ , \qquad m_A^2 \sqrt{\lambda_B} = m_B^2 \sqrt{\lambda_A} \ . \tag{19}$$

In this case the values of the two minima in (15) and (16) are degenerate and there is a *minimum valley* connecting them along the line

$$(A^r B^r)^2 = A^r A^r B^s B^s \ , \qquad A^r A^r \sqrt{\lambda_A} + B^r B^r \sqrt{\lambda_B} = -\frac{m_A^2}{4\sqrt{\lambda_A}} = -\frac{m_B^2}{4\sqrt{\lambda_B}} \ . \tag{20}$$

If one transforms the vacuum expectation values, say, to the third direction by a global gauge transformation, then in the $(A^3, B^3)$ plane this defines an ellipse. For $\lambda_A = \lambda_B$ the ellipse becomes a circle.

The quantum corrections are, in general, changing the positions of the minima and the values of the effective potential at the minima. It can, however, be expected that at least for weak couplings the phase structure remains qualitatively the same. Particularly interesting is the fate of the minimum valley defined by (20) because of the non-renormalization theorems, which tell that the ambiguity of the minimum position is not resolved in a supersymmetric situation (see [8, 7]).

# 3 Tuning the parameters

In this section a way of approaching the supersymmetric continuum limit will be proposed. Only the Higgs phases will be considered here because they are more interesting physically. Since the $N = 2$ supersymmetric Yang-Mills theory is asymptotically free, the perturbative $\beta$-functions can give qualitative information about the behaviour of the *lines of constant physics* (LCPs) in bare parameter space. These lines, which are defined by keeping the independent renormalized couplings fixed, go to the continuum limit, if it exists. The naïve prescription to simply look for the LCP in bare parameter space, which corresponds to the supersymmetric relations in (8), does not work because of the conflict with the path integral stability conditions (12).

This means that one has to consider the renormalization group equations (RGEs) of all renormalizable couplings, including also those which are not asymptotically free. The two-loop RGE for the gauge coupling is:

$$\frac{dg}{dt} = -\frac{4g^3}{(4\pi)^2} + \frac{8g^3}{(4\pi)^4}(2g^2 - G_A^2 - G_B^2) + \cdots . \quad (21)$$

For the other couplings [9, 10] one can give the equations immediately in terms of the ratios

$$R_{A,B} \equiv \frac{G_{A,B}}{g} , \qquad r_{A,B} \equiv \frac{\lambda_{A,B}}{g^2} , \qquad r_{[AB]} \equiv \frac{\lambda_{[AB]}}{g^2} , \qquad r_{(AB)} \equiv \frac{\lambda_{(AB)}}{g^2} . \quad (22)$$

Up to one-loop order we have

$$\frac{dR_A}{dt} = \frac{g^2}{(4\pi)^2} \cdot 8R_A(R_A^2 - 1) + \mathcal{O}\left[\frac{g^4}{(4\pi)^4}\right] ,$$

$$\frac{dR_B}{dt} = \frac{g^2}{(4\pi)^2} \cdot 8R_B(R_B^2 - 1) + \mathcal{O}\left[\frac{g^4}{(4\pi)^4}\right] ,$$

$$\frac{dr_A}{dt} = \frac{g^2}{(4\pi)^2} \left[3 - 16r_A + 16r_A R_A^2 - 4R_A^4 + 88r_A^2 + 2r_{(AB)}^2 + 6r_{[AB]}^2 - 4r_{(AB)}r_{[AB]}\right] + \mathcal{O}\left[\frac{g^4}{(4\pi)^4}\right] ,$$

$$\frac{dr_B}{dt} = \frac{g^2}{(4\pi)^2} \left[3 - 16r_B + 16r_B R_B^2 - 4R_B^4 + 88r_B^2 + 2r_{(AB)}^2 + 6r_{[AB]}^2 - 4r_{(AB)}r_{[AB]}\right] + \mathcal{O}\left[\frac{g^4}{(4\pi)^4}\right] ,$$

$$\frac{dr_{[AB]}}{dt} = \frac{g^2}{(4\pi)^2} \Big[3 - 16r_{[AB]} + 8r_{[AB]}(R_A^2 + R_B^2) - 8R_A^2 R_B^2$$

$$+ 8(r_A + r_B)(5r_{[AB]} - r_{(AB)}) + 16r_{[AB]}^2 + 4r_{(AB)}^2\Big] + \mathcal{O}\left[\frac{g^4}{(4\pi)^4}\right] ,$$

$$\frac{dr_{(AB)}}{dt} = \frac{g^2}{(4\pi)^2} \Big[-3 - 16r_{(AB)} + 8r_{(AB)}(R_A^2 + R_B^2)$$

$$+ 16(r_A + r_B)r_{(AB)} - 20r_{(AB)}^2 + 32r_{(AB)}r_{[AB]}\Big] + \mathcal{O}\left[\frac{g^4}{(4\pi)^4}\right] . \quad (23)$$

Here the right-hand sides vanish at the supersymmetric point

$$\mathbf{r}_{susy} \equiv \left( R_A = 1, \ R_B = 1, \ r_A = 0, \ r_B = 0, \ r_{[AB]} = \frac{1}{2}, \ r_{(AB)} = \frac{1}{2} \right) . \tag{24}$$

In other words, supersymmetry corresponds to a *fixed point* of the coupling ratios, which we shall call in the following simply the supersymmetric fixed point. The stability properties of this fixed point play a crucial rôle in parameter tuning for the supersymmetric continuum limit.

The RGEs in eqs. (21) and (23) hold, for instance, for the bare couplings at fixed renormalized couplings. This means that they give the LCPs in bare parameter space. In this case $t = \log(m_{physical}^{-1})$, where $m_{physical}$ is some physical mass in lattice units. The continuum limit corresponds to $t \to \infty$. The same one-loop equations also describe the change of the renormalized couplings for fixed bare couplings, if the right-hand sides are multiplied by overall minus signs. This means that in the latter case the flow of couplings is reversed: what was an attractive fixed point before becomes a repulsive one and vice versa.

The stability of the supersymmetric fixed point can be investigated in the linearized approximation of eqs. (23) near $\mathbf{r} = \mathbf{r}_{susy}$:

$$\frac{d\mathbf{r}}{dt} = \frac{g^2}{(4\pi)^2} \cdot \mathbf{D}(\mathbf{r} - \mathbf{r}_{susy}) + \cdots . \tag{25}$$

The matrix $\mathbf{D}$ is given by

$$\mathbf{D} = \begin{pmatrix} 16 & 0 & 0 & 0 & 0 & 0 \\ 0 & 16 & 0 & 0 & 0 & 0 \\ -16 & 0 & 0 & 0 & 4 & 0 \\ 0 & -16 & 0 & 0 & 4 & 0 \\ -8 & -8 & 16 & 16 & 16 & 4 \\ 8 & 8 & 8 & 8 & 16 & -4 \end{pmatrix} . \tag{26}$$

The eigenvalues $\alpha_{1...6}$ and eigenvectors $\mathbf{e}_{1...6}$ of this matrix are the following:

$$\begin{aligned}
\alpha_1 &= 0 \ , & \mathbf{e}_1 &= (0,0,1,-1,0,0) \ , \\
\alpha_2 &= 16 \ , & \mathbf{e}_2 &= (1,1,0,0,4,4) \ , \\
\alpha_3 &= 16 \ , & \mathbf{e}_3 &= (2,0,-1,1,4,4) \ , \\
\alpha_4 &= -8 \ , & \mathbf{e}_4 &= (0,0,1,1,-2,4) \ , \\
\alpha_5 &= -4 \ , & \mathbf{e}_5 &= (0,0,1,1,-1,-3) \ , \\
\alpha_6 &= 24 \ , & \mathbf{e}_6 &= (0,0,1,1,6,4) \ .
\end{aligned} \tag{27}$$

Since there are eigenvalues with different signs, the supersymmetric fixed point is neither attractive nor repulsive: its behaviour depends on the direction of approaching it. Three of the eigenvectors, namely $\mathbf{e}_{1,3,4}$, point towards directions where the conditions for path integral stability in (12) are violated. They have to be excluded by appropriate parameter tuning. Two out of the remaining three directions ($\mathbf{e}_{2,6}$) have positive eigenvalues and there is only a single direction ($\mathbf{e}_5$) with negative eigenvalue. A consequence of this is that the supersymmetric continuum limit at $(g^2 = 0, \mathbf{r} = \mathbf{r}_{susy})$ can only be reached by an LCP if it comes from this unique direction in bare coupling space. All other LCPs go either to infinity in some component of **r**, or end by violating eq. (12). The LCPs reaching a supersymmetric continuum limit are the *ways to supersymmetry*.

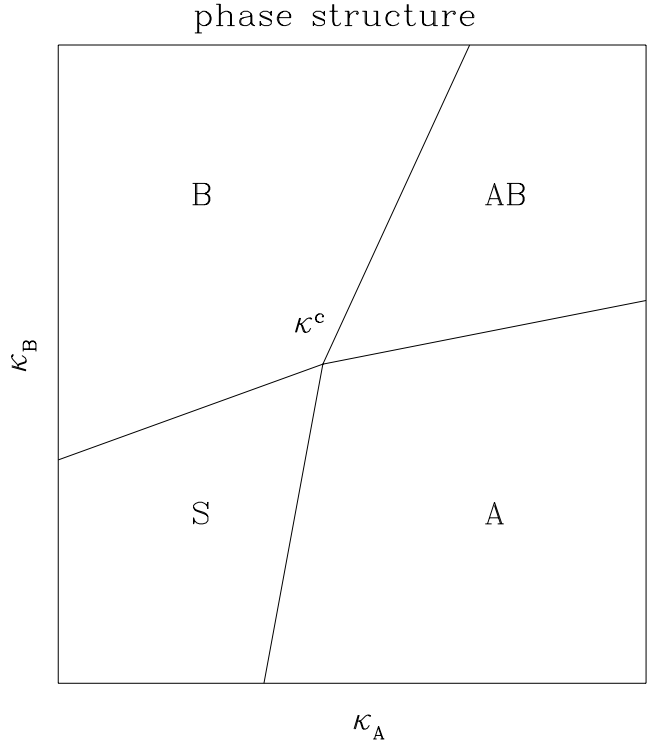

Figure 1: The schematic view of the generic phase structure in the plane of scalar hopping parameters near the critical point $\kappa^c$. Besides $\kappa_{A,B}$ all other bare parameters are fixed; $S$ is the symmetric phase, $A$ and $B$ correspond to non-zero vacuum expectation values of the $A$- and $B$-fields, respectively. In the phase $AB$ both scalar fields have non-zero vacuum expectation values.

The remaining question is the tuning of the bare mass parameters $(m_\psi, m_A, m_B)$. In a numerical simulation they are usually represented by the corresponding hopping parameters $(\kappa_\psi, \kappa_A, \kappa_B)$. For fixed bare couplings ($g^2$ and **r**) one expects the existence of a critical line $\kappa^c(\kappa_\psi) \equiv (\kappa_A = \kappa_A(\kappa_\psi), \kappa_B = \kappa_B(\kappa_\psi))$ where all scalar boson masses are zero in lattice units. The existence of such a line of critical points, corresponding to a second-order phase transition, has to be checked by numerical simulations. The condition for tuning the fermion hopping parameter $\kappa_\psi$ is the restoration of the global $SU(2)_\mathcal{R}$ symmetry. This happens if the diagonal part of the fermion mass matrix (in triplet index) vanishes. Denoting this value of $\kappa_\psi$ by $\kappa_\psi^c$, the remaining task is to tune the scalar hopping parameters $\kappa_{A,B}$ to the critical point $\kappa^c \equiv (\kappa_\psi^c, \kappa_A^c, \kappa_B^c)$. Going to some other point of the critical line $\kappa^c(\kappa_\psi)$, the result is supersymmetry softly broken by a Dirac fermion mass.

On the basis of the analysis of the minima of the classical potential the generic phase structure near $\kappa^c$ is expected to look as is shown in fig. 1. In sector $AB$ the relevant minimum of the effective potential corresponds to the minimum in eq. (18). As has been discussed there, for $4\lambda_A\lambda_B = (\lambda_{[AB]} - \lambda_{(AB)})^2$ this sector is shrinking to a line defined by (19). This refers, however, only to the classical theory. In the case of interest for supersymmetry, when $\lambda_A = \lambda_B$, the points of the minimum valley defined in (20) can be parametrized by the global chiral transformation of the scalar fields given in (10). This symmetry is, however, anomalous; therefore it has to be explicitly broken in lattice regularization. In fact, the Wilson term proportional to $r$ in the lattice action (3) breaks it and the anomaly is emerging in the continuum limit in the well-known way [11]. Since the chiral symmetry is broken in the fermion sector, the degeneracy of the minima in the valley is removed by quantum corrections and the $AB$ phase never collapses to a line. Therefore the ratio of the vacuum expectation values of the $A$- and $B$-fields can be arbitrarily tuned by the angle of approaching the critical point $\kappa^c$ in the $(\kappa_A, \kappa_B)$-plane. This ratio is a relevant parameter of different supersymmetric continuum limits. Before discussing the tuning of the overall lengths of the vacuum expectation values let us return to the RGEs in (21), (23).

As stated before, the RGEs with a minus sign on the right-hand sides can also be used, for sufficiently small couplings, to describe the change of the renormalized couplings for fixed bare couplings. Since the signs of the eigenvalues in (27) are now reversed, there are two attractive directions: $\mathbf{e}_2$ and $\mathbf{e}_6$. Tuning the initial values of the renormalized couplings with the help of the bare couplings to the plane spanned by these two vectors, and then going towards the critical point for fixed bare couplings, the ratios of renormalized couplings are approaching the supersymmetric relations in (8).

The overall ratios of the renormalized vacuum expectation values of the $A$- and $B$-fields to the $\Lambda$-parameter of the gauge coupling $v_{Ar}/\Lambda_{LATT}$, $v_{Br}/\Lambda_{LATT}$ can be tuned as follows. Proceeding as described in the previous paragraph and assuming that one starts from a region where the scalar masses are of order 1 in lattice units, the initial values of these ratios are essentially fixed by the initial value of the bare gauge coupling $g^2$. Since $\Lambda_{LATT} \propto \exp(-2\pi^2/g^2)$ becomes smaller for decreasing $g^2$, the ratios $v_{Ar}/\Lambda_{LATT}$, $v_{Br}/\Lambda_{LATT}$ become larger. That is for large $\beta = 4g^{-2}$ one obtains large vacuum expectation values in units of the $\Lambda$-parameter of the gauge coupling. Note that, in contrast to pure gauge theory or QCD, in Higgs phases the numerical simulation at large $\beta$ is not a problem [12].

In summary, the renormalized vacuum expectation values of the scalar fields can be arbitrarily tuned by the two scalar hopping parameters. Therefore the points of the moduli space of quantum vacua investigated in ref. [1] correspond to different physical theories, much as, for instance, QCD with different values of the quark masses.

## 4  Outlook

Tuning the parameters in the SU(2) adjoint Higgs-Yukawa model on the lattice to reach the $N = 2$ supersymmetric continuum limit can be guided by the fixed point properties of the renormalization group equations. In particular, one can combine the use of a single attractive direction for bare couplings and two linearly independent attractive directions for renormalized couplings. For the successful tuning of bare parameters of the lattice theory to reach super-symmetry in the continuum limit a decisive question is the non-perturbative phase structure (see fig. 1). If the appropriate phase structure is reproduced, the points of the quantum moduli space of vacua are the endpoints of a two-parameter family of LCPs, which represent all possible ways to $N = 2$ supersymmetry in the SU(2) adjoint Higgs-Yukawa model.

The lattice methods are well suited for these kinds of investigations. There is extensive experience in simulating SU(2) Higgs and Yukawa models with scalar and fermion doublets. (For the more difficult case of Yukawa models see e.g. [13].) The replacement of doublets by triplets in the adjoint representation does not seem to be a problem.

The crucial question, which finally will decide whether the numerical simulation can be performed by presently known methods, is the behaviour and implementation of the fermionic determinant. In several popular and effective simulation algorithms, such as the Hybrid Monte Carlo, the number of fermion fields has to be doubled, which would destroy the balance between fermionic and bosonic fields. This is, of course, prohibited in a supersymmetric continuum limit. There are algorithms, such as the hybrid classical-Langevin algorithm (see e.g. [14]) where the appropriate number of fermions can be simulated, but only under the condition that the phase of the fermionic determinant is negligible. Taking into account small fluctuations of the phase is possible by the method developed in [15]. Large phase fluctuations are, however, at present not tolerable.

In case of the SU(2) adjoint Higgs-Yukawa model the fermion determinant is real therefore the phase factor can only be $\pm 1$. Previous experience with fermion determinants in other models makes it plausible that the fluctuation of this sign is not important. In any case, this can be checked by explicitly calculating the fermion determinant [15]. If the effect of the sign of the fermion determinant is small indeed, then the numerical simulation of $N = 2$ supersymmetry becomes possible. I hope to return to this question in a future publication.

## Acknowledgement

It is a pleasure to thank Gabriele Veneziano, Luis Alvarez-Gaumé and Rainer Sommer for helpful and enlightening discussions.